\documentclass[aps,prx,twocolumn,showpacs,superscriptaddress]{revtex4-1}
\usepackage{bm}
\usepackage{mathrsfs}
\usepackage{amsmath}
\usepackage{amssymb}
\usepackage{graphicx}
\usepackage{amsfonts}
\usepackage{amsthm}
\usepackage{color}
\usepackage{dcolumn}
\usepackage{txfonts}
\usepackage{epsfig}

\begin{document}

\title{Links between Dissipation and R\'{e}nyi Divergences in $\mathcal{PT}$-Symmetric Quantum Mechanics}
\author{Bo-Bo Wei}
\affiliation{School of Physics and Energy, Shenzhen University, 518060 Shenzhen, China}

\begin{abstract}
Thermodynamics and information theory have been intimately related since the times of Maxwell and Boltzmann. Recently it was shown that the dissipated work in an arbitrary non-equilibrium process is related to the R\'{e}nyi divergences between two states along the forward and reversed dynamics. Here we show that the relation between dissipated work and Renyi divergences generalizes to $\mathcal{PT}$-symmetric quantum mechanics with unbroken $\mathcal{PT}$ symmetry. In the regime of broken $\mathcal{PT}$ symmetry, the relation between dissipated work and Renyi divergences do not hold as
the norm is not preserved during the dynamics. This finding is illustrated for an
experimentally relevant system of two-coupled cavities.
\end{abstract}
\maketitle

\section{Introduction}
In 1997, Jarzynski \cite{Jarzynski1997} discovered that for a classical system initialized in a thermodynamic equilibrium state the work done under an arbitrary change of control parameters is related to the equilibrium free energy difference between the initial and the final thermal equilibrium states for the control parameters,
\begin{eqnarray}\label{JE}
\langle e^{-\beta W}\rangle=\frac{Z(\beta,\lambda_f)}{Z(\beta,\lambda_i)}=e^{-\beta [F(\beta,\lambda_f)-F(\beta,\lambda_i)]}.
\end{eqnarray}
Here $\beta\equiv 1/T$ is the inverse temperature of the initial thermodynamic equilibrium state of the classical system and the Boltzmann constant is set to unity, $k_B\equiv1$, $W$ is the work done on the system in a driving process which varies the control parameter $\lambda$ from $\lambda_i$ to $\lambda_f$, $F\equiv-\beta^{-1}\ln Z$ is the Helmholtz free energy of the system with $Z$ being the equilibrium partition function and the angular bracket on the left of Equation \eqref{JE} denotes an ensemble average over all realizations of the driving process. The Jarzynski equality bridges the equilibrium free energy difference, and a non-equilibrium quantity, the work done in an arbitrary driving process and therefore we can measure the equilibrium free energy difference of a classical system by repeatedly performing work on the system by any protocol. Later, Jarzynski equality was generalized to finite quantum mechanical systems \cite{arXiv2000a,arXiv2000b,PRL2003,Talker2007} and has been verified experimentally in various physical systems \cite{PNAS2001,Science2005,Nature2005,EPL2005,PT2005,PRL2006,PRL2007,PRL2012,Kim2015}. The discovery of the Jarzynski equality has led to an active investigation of various fluctuation relations in non-equlibrium thermodynamics \cite{RMP2009,Jar2011,RMP2011}.

Recently, the author and his collaborator found that \cite{Wei2017a} the dissipated work $W_{\text{diss}}\equiv W-\Delta F$ satisfies a neat and exact
microscopic fluctuation relation,
\begin{eqnarray}\label{central}
\langle \Big(e^{-\beta W_{\text{diss}}}\Big)^z\rangle =e^{(z-1)S_{z}[\Theta\rho_R(\tau-t)\Theta^{-1}||\rho(t)]}.
\end{eqnarray}
Here $z$ is a finite real number, $W_{\text{diss}}$ is the dissipated work done on the system in a driving protocol under which the control parameter varies from $\lambda_i$ at $t=0$ to $\lambda_f$ in time $\tau$, the angular bracket on the left hand side denotes an ensemble average over the realizations of the driving process and $S_{z}[\Theta\rho_R(\tau-t)\Theta^{-1}||\rho(t)]$ is the order-$z$ R\'{e}nyi divergence between a non-equilibrium state in the forward process $\rho(t)$ and a non-equilibrium state in its time reversed process $\Theta\rho_R(\tau-t)\Theta^{-1}$ with $\Theta$ being the time reversal operator. For classical system, the order-$z$ R\'{e}nyi divergence between two distributions $\rho_1(X)$ and $\rho_2(X)$ is defined as \cite{Renyi1961,Erven2014}  $S_{z}[\rho_1||\rho_2]\equiv\frac{1}{z-1}\ln[\int dX\rho_1^{z}(X)\rho_2^{1-z}(X)]]$. For quantum system, the order-$z$ R\'{e}nyi divergence betwen quantum states $\rho_1$ and $\rho_2$ is defined as \cite{Beigi2013,Lennert2013} $S_{z}[\rho_1||\rho_2]\equiv\frac{1}{z-1}\ln[\text{Tr}[\rho_1^{z}\rho_2^{1-z}]]$. Equation \eqref{central} can be considered as a generalization of Jarzynski equality. On the one hand, the
relation in Equation \eqref{central} shows that the macroscopic entropy production and its fluctuations in a non-equilibrium process are determined by
the family of R\'{e}nyi divergences which are a measure of distinguishability of two states \cite{Beigi2013,Lennert2013} between a microscopic
process and its time reversed process. On the other hand, the relation in Equation \eqref{central} tells us that we can measure the family of
R\'{e}nyi divergences by non-equilibrium work measurement in a microscopic process.

Recently the quantum Jarzynski equality was extended to the $\mathcal{PT}$-symmetric quantum systems \cite{PT2015,PT2016,PT2017}. The motivation of the present work is to investigate whether the links between dissipated work and R\'{e}nyi divergences survives in $\mathcal{PT}$ symmetric quantum systems? We found that this is indeed the case for systems in the phase of unbroken $\mathcal{PT}$ symmetry. On the other hand, for systems in the phase of broken $\mathcal{PT}$ symmetry, the relation between dissipated work and R\'{e}nyi divergences does not hold.

The remainder of this paper is organized as follows: In Sec.~II, we review the formalism of $\mathcal{PT}$-symmetric quantum mechanics. In Sec.~III, we first review the formalism for quantum thermodynamics for $\mathcal{PT}$-symmetric quantum systems and then derive the relations between the dissipated work in driving $\mathcal{PT}$-symmetric quantum system and the family of R\'{e}nyi divergences between two quantum states. In Sec. IV, we use an experimentally relevant $\mathcal{PT}$-symmetric quantum systems to verify our finding. In Sec.~V, we make a brief summary.

\section{Fundamentals of $\mathcal{PT}$-symmetric Quantum Mechanics}
We first review the key properties of $\mathcal{PT}$-symmetric quantum mechanics. For a non-Hermitian Hamiltonian $\mathcal{H}$ with $\mathcal{PT}$ symmetry, it is well known that the Hamiltonian $\mathcal{H}$ generally presents two different phases, namely, the phase of unbroken $\mathcal{PT}$ symmetry in which all the eigenvalues are real and the phase of broken $\mathcal{PT}$ symmetry for which the eigenvalue spectrum has real and imaginary parts. The universal critical behavior for phase transitions in such non-Hermitian Hamiltonian has been revealed recently \cite{Wei2017b}.

For non-Hermitian Hamiltonian \cite{NonHermitian2011}, $\mathcal{H}\neq\mathcal{H}^{\dagger}$, the left and right eigenvectors are different, namely
\begin{eqnarray}
\mathcal{H}|\psi_n\rangle&=&E_n|\psi_n\rangle,\\
\mathcal{H}^{\dagger}|\phi_n\rangle&=&E_n^*|\phi_n\rangle.
\end{eqnarray}
Here for simplicity of notation, we assume the eigenstates are discrete and $\langle\psi_n|\phi_m\rangle=\delta_{mn}$ and $\sum_n|\psi_n\rangle\langle\phi_n|=1$. Existence of biorthonormal sets of eigenvectors means that \cite{NonHermitian2011}
\begin{eqnarray}
\mathcal{H}^{\dagger}=\mathcal{G}\mathcal{H}\mathcal{G}^{-1}.
\end{eqnarray}
Here $\mathcal{G}$ and $\mathcal{G}^{-1}$ are given respectively by \cite{NonHermitian2011}
\begin{eqnarray}
\mathcal{G}&=&\sum_n|\phi_n\rangle\langle\phi_n|,\\
\mathcal{G}^{-1}&=&\sum_n|\psi_n\rangle\langle\psi_n|.
\end{eqnarray}
Thus the normalization condition for arbitrary state $|\Psi\rangle$ is given by \cite{NonHermitian2011}
\begin{eqnarray}
\langle\Psi|\mathcal{G}|\Psi\rangle=1.
\end{eqnarray}
The completeness relation is \cite{NonHermitian2011}
\begin{eqnarray}
\sum_m|\phi_m\rangle\langle\phi_m|\mathcal{G}=1.
\end{eqnarray}

For a time-dependent Hamiltonian $\mathcal{H}(t)$ with $\mathcal{PT}$ symmetry, the dynamics of a state is still governed by the Schr\"{o}dinger equation but a modification has to be made to preserve unitarity. In this framework, the modified Schr\"{o}dinger equation is given by \cite{PTevolution}
\begin{eqnarray}
i\partial_t|\psi\rangle=[\mathcal{H}(t)+\mathcal{A}(t)]|\psi\rangle,
\end{eqnarray}
where $\mathcal{A}(t)=-i\mathcal{G}_t^{-1}\partial_t\mathcal{G}_t$ is a time-dependent gauge field that has been added to guarantee unitarity of the non-Hermitian quantum dynamics when all the energy eigenvalues are real. The corresponding time evolution operator can be written as
\begin{eqnarray}\label{evolution}
\mathcal{U}_{0,t}=\mathcal{T}e^{-i\int_0^tdt'[\mathcal{H}(t')+\mathcal{A}(t')]}.
\end{eqnarray}
Here $\mathcal{U}_{0,t}$ is the time evolution operator from time $0$ to $t$ and $\mathcal{T}$ is the time ordering operator which comes from the time-dependent Hamiltonian in the Schr\"{o}dinger equation. For the quantum system in the unbroken $\mathcal{PT}$-symmetric phase, all the eigenvalues are real and thus the quantum dynamics generated by $\mathcal{U}_{0,t}$ is unitary in the sense that the probability is conserved. However, the time evolution operator $\mathcal{U}_{0,t}$ is not unitary but satisfies
\begin{eqnarray}\label{unitarity}
\mathcal{U}_{0,t}^{\dagger}\mathcal{G}_{t}\mathcal{U}_{0,t}=\mathcal{G}_0.
\end{eqnarray}
This relation is shown in Appendix A and it can be considered as the corresponding unitarity condition in $\mathcal{PT}$-symmetric quantum mechanics and it reduces back to the unitarity condition for Hermitian quantum mechanics when $\mathcal{G}\equiv I$.

To tackle problems in quantum thermodynamics of $\mathcal{PT}$-symmetric systems,  some basic mathematical operations have to be modified to match the theory of statistical mechanics. The key modification is the form of inner product. In the non-Hermitian formalism the inner product has to be modified to $\langle\phi|\psi\rangle\rightarrow\langle\phi|\mathcal{G}|\psi\rangle$. Correspondingly, the trace operation has to be changed as \cite{NonHermitian2011,PTevolution}
\begin{eqnarray}
\text{Tr}_{\mathcal{G}}[\mathcal{M}]=\sum_n\langle\psi_n|\mathcal{GM}|\psi_n\rangle.
\end{eqnarray}
Here $\mathcal{M}$ is some arbitrary operator and $\{|\psi_n\rangle\}$ form a complete basis in the Hilbert space. It is easy to show that this trace operation satisfies the cyclic property \cite{PT2017}, namely $\text{Tr}_{\mathcal{G}}[\mathcal{AB}]=\text{Tr}_{\mathcal{G}}[\mathcal{BA}]$. Recently thermodynamics with a complex control parameter has been extensively investigated by the quantum decoherence
\cite{Wei2012,Wei2014,Wei2015,Peng2015,LYExp2015,Wei2017a1,Wei2017b1} and large derivation statistics \cite{DynamicLY2013,DynamicLY2017}.

\section{Non-equilibrium Quantum thermodynamics for $\mathcal{PT}$-symmetric Systems}
Let us consider a finite quantum system with $\mathcal{PT}$ symmetry and we are interested in a non-equilibrium process induced by a time-dependent parameter $\lambda(t)$ which is controlled by an external agent. We first consider a driving protocol for which the Hamiltonian $\mathcal{H}(t)\equiv\mathcal{H}(\lambda(t))$ is in the phase of unbroken $\mathcal{PT}$ symmetry at all times.

Let us first define the \emph{non-equilibrium process in $\mathcal{PT}$-symmetric quantum system} under time-dependent driving. We initialize the quantum system in canonical equilibrium state at inverse temperature $\beta=1/T$ at a fixed value of control parameter $\lambda_i$, which is described by the density matrix
\begin{eqnarray}
\rho(0)=e^{-\beta \mathcal{H}(\lambda_i)}/Z(\beta,\lambda_i),
\end{eqnarray}
where $Z(\beta,\lambda_i)=\text{Tr}_{\mathcal{G}}[e^{-\beta \mathcal{H}(\lambda_i)}]$ is the partition function. Then we isolate the system and drive it by the Hamiltonian $\mathcal{H}(\lambda(t))$ for a time duration $\tau$, where the force protocol $\lambda(t), t\in[0,\tau]$ brings the parameter from $\lambda_i$ at $t=0$ to $\lambda_f$ at a later time $\tau$. Then the state at $t$ in the driving process is given by
\begin{eqnarray}\label{ef}
\rho(t)&=&\mathcal{U}_{0,t}\rho(0)\mathcal{U}_{0,t}^{-1},
\end{eqnarray}
where $\mathcal{U}_{0,t}\equiv \mathcal{T}e^{-i\int_0^tdt'[\mathcal{H}(\lambda(t'))+\mathcal{A}(t')]}$ with $\mathcal{T}$ being the time ordering operator. Equation \eqref{ef} is derived in the Appendix B.  Generally, the final non-equilibrium state $\rho(\tau)$ is different from the thermodynamic equilibrium state at the final control parameter, namely, $\rho_f=e^{-\beta \mathcal{H}(\lambda_f)}/Z(\beta,\lambda_f)$,  which is only achievable by adiabatic process.

It is generally accepted that work in quantum system is defined by two projective measurements \cite{Talker2007,RMP2011}.
We assume, for any control parameter $\lambda$, $\mathcal{H}(\lambda)|n(\lambda)\rangle=E_n(\lambda)|n(\lambda)\rangle$ and the quantum number $n$ labels eigen vectors. At $t=0$, we perform the first projective measurement of $\mathcal{H}(\lambda_i)$ and the outcome is $E_n(\lambda_i)$ with probability,
\begin{eqnarray}
p_n(0)=\text{Tr}_{\mathcal{G}_0}[\rho(0)|n(\lambda_i)\rangle\langle n(\lambda_i)|\mathcal{G}_0]=e^{-\beta E_n(\lambda_i)}/Z(\beta,\lambda_i).
\end{eqnarray}
Simultaneously the initial equilibrium state $\rho(0)$ projects into the corresponding eigen state $|n(\lambda_i)\rangle$. In the time interval, $0<t<\tau$, the quantum system is isolated and driven by the evolution operator $\mathcal{U}_{0,\tau}$ and then the quantum state at time $\tau$ is $\mathcal{U}_{0,\tau}|n(\lambda_i)\rangle$. Finally at time $t=\tau$, we perform the second projective measurement of $\mathcal{H}(\lambda_f)$ and the result is $E_m(\lambda_f)$ with conditional probability,
\begin{eqnarray}
p_{n\rightarrow m}=|\langle m(\lambda_f)|\mathcal{G}_{\tau}\mathcal{U}_{0,\tau}|n(\lambda_i)\rangle|^2.
\end{eqnarray}
Summarizing the two projective measurements of energy, the probability of obtaining $E_n(\lambda_i)$ for the first measurement and followed by getting $E_m(\lambda_f)$ in the second measurement is $p_n(0)p_{n\rightarrow m}$. Hence the quantum work distribution is given by \cite{Talker2007,RMP2011}
\begin{eqnarray}\label{QWD}
P(W)=\sum_{m,n}p_n(0)p_{n\rightarrow m}\delta\left(W-E_m(\lambda_f)+E_n(\lambda_i)\right).
\end{eqnarray}
The quantum work distribution $P(W)$ encodes the fluctuations in the work
that arise from thermal statistics and from quantum
measurement statistics over many identical realizations
of the protocol. The characteristic function of quantum work distribution is given by \cite{Talker2007,RMP2011}
\begin{eqnarray}\label{charac}
G(u)&=&\int_{-\infty}^{\infty}dWP(W)e^{iuW},\\
&=&Z(\beta,\lambda_i)^{-1}\text{Tr}_{\mathcal{G}_0}[\mathcal{U}_{0,\tau}e^{-(\beta+iu)\mathcal{H}(0)}\mathcal{U}_{0,\tau}^{-1}e^{iu\mathcal{H}(\tau)}].
\end{eqnarray}

Now let us define the reversed process in quantum system with $\mathcal{PT}$ symmetry. In the reversed process, we first initialize the system in the time reversal of the thermodynamic equilibrium state at inverse temperature $\beta=1/T$ at control parameter $\lambda_f$,
\begin{eqnarray}
\rho_R(0)=\frac{\Theta e^{-\beta \mathcal{H}(\lambda_f)}\Theta^{-1}}{Z(\beta,\lambda_f)},
\end{eqnarray}
where $Z(\beta,\lambda_f)=\text{Tr}_{\mathcal{G}}[e^{-\beta\mathcal{H}(\lambda_f)}]$ is the canonical
partition function. Then we drive the system in the reversed process for a time interval $\tau$ by the Hamiltonian \cite{Stra1994},
\begin{eqnarray}
\mathcal{H}_R(t)=\Theta \mathcal{H}(\tau-t)\Theta^{-1}.
\end{eqnarray}
The time evolution operator in the reversed process is then given by
\begin{eqnarray}
\mathcal{V}_{0,\tau}=\mathcal{T}e^{-i\int_0^{\tau}dt[\Theta(\mathcal{H}(\tau-t)+\mathcal{A}(\tau-t))\Theta^{-1}]}.
\end{eqnarray}
So the evolution of the density operator in the reversed process is
\begin{eqnarray}\label{er}
\rho_R(t)&=&\mathcal{V}_{0,t}\rho_R(0)\mathcal{V}_{0,t}^{-1}.
\end{eqnarray}
The time evolution operator in the forward process and its time reversed process are related by (See Appendix C for proof)
\begin{eqnarray}\label{frrelation}
\mathcal{V}_{0,t}=\Theta\mathcal{U}_{\tau-t,\tau}^{-1}\Theta^{-1}.
\end{eqnarray}
Although $\rho(t)$ and $\rho_R(t)$ are far from equilibrium states, they satisfy the following lemma due to time reversal
symmetry and unitarity of the time development:\\
\textbf{Lemma}. The density operator $\rho(t)$ in the forward driving process at arbitrary time $t\in[0,\tau]$ and the density operator $\rho_R(\tau-t)$ in the time reversed process
at time $\tau-t$ satisfy, for any finite real numbers $a,b\in\mathcal{R}$,
\begin{eqnarray}
\text{Tr}_{\mathcal{G}}[\left(\Theta^{-1}\rho_R(\tau-t)\Theta\right)^a\rho(t)^b]=\text{Tr}_{\mathcal{G}}[\left(\Theta^{-1}\rho_R(\tau)\Theta\right)^a\rho(0)^b].\nonumber\\
\end{eqnarray}
\textbf{Proof:} By Equation \eqref{ef}, we have
\begin{eqnarray}
\Theta^{-1}\rho_R(\tau-t)\Theta&=&\Theta^{-1}\mathcal{V}_{0,\tau-t}\rho_R(0)\mathcal{V}_{0,\tau-t}^{-1}\Theta.
\end{eqnarray}
Then
\begin{eqnarray}
&&\left(\Theta^{-1}\rho_R(\tau-t)\Theta\right)^a\rho(t)^b,\\
&=&\left(\Theta^{-1}\mathcal{V}_{0,\tau-t}\rho_R(0)\mathcal{V}_{0,\tau-t}^{-1}\Theta\right)^a\left(\mathcal{U}_{0,t}\rho(0)\mathcal{U}_{0,t}^{-1}\right)^b,\\
&=&\Theta^{-1}\mathcal{V}_{0,\tau-t}\rho_R(0)^a\mathcal{V}_{0,\tau-t}^{-1}\Theta\mathcal{U}_{0,t}\rho(0)^b\mathcal{U}_{0,t}^{-1},\\
&=&\Theta^{-1}\mathcal{V}_{\tau-t,\tau}^{-1}\mathcal{V}_{0,\tau}\rho_R(0)^a\mathcal{V}_{0,\tau}^{-1}\mathcal{V}_{\tau-t,\tau}^{-1}\Theta\nonumber
\\&&\times\Theta^{-1}\mathcal{V}_{0,\tau-t}\mathcal{V}_{0,\tau}^{-1}\Theta\rho(0)^b
\Theta^{-1}\mathcal{V}_{0,\tau}\mathcal{V}_{0,\tau-t}^{-1}\Theta,\\
&=&\Theta^{-1}\mathcal{V}_{\tau-t,\tau}^{-1}\mathcal{V}_{0,\tau}\rho_R(0)^a\mathcal{V}_{0,\tau}^{-1}\Theta\rho(0)^b
\Theta^{-1}\mathcal{V}_{\tau-t,\tau}\Theta,\\
&=&\Theta^{-1}\mathcal{V}_{\tau-t,\tau}^{-1}\rho_R(\tau)^a\Theta\rho(0)^b
\Theta^{-1}\mathcal{V}_{\tau-t,\tau}\Theta.
\end{eqnarray}
Taking trace on the both sides of the above equation and applying the cyclic invariance of the trace operation, we get
\begin{eqnarray}
&&\text{Tr}_{\mathcal{G}}[\left(\Theta^{-1}\rho_R(\tau-t)\Theta\right)^a\rho(t)^b],\\
&=&\text{Tr}_{\mathcal{G}}[\Theta^{-1}\mathcal{V}_{\tau-t,\tau}^{-1}\rho_R(\tau)^a\Theta\rho(0)^b
\Theta^{-1}\mathcal{V}_{\tau-t,\tau}\Theta],\\
&=&\text{Tr}_{\mathcal{G}}[\Theta^{-1}\rho_R(\tau)^a\Theta\rho(0)^b],\\
&=&\text{Tr}_{\mathcal{G}}[\left(\Theta^{-1}\rho_R(\tau)\Theta\right)^a\rho(0)^b].
\end{eqnarray}
Thus the Lemma is proved.

From the definition of quantum work distribution, Equation\eqref{QWD}, we have
\begin{eqnarray}
&&\langle\Big( e^{-\beta W}\Big)^z\rangle=\int dW P(W)e^{-\beta zW},\\
&=&Z(\lambda_i)^{-1}\sum_{n}\langle n(\lambda_i)|\mathcal{U}_{0,\tau}^{\dagger}\mathcal{G}_{\tau}e^{-\beta z\mathcal{H}(\tau)}\mathcal{U}_{0,\tau}e^{-\beta(1-z) \mathcal{H}(0)}|n(\lambda_i)\rangle,\\
&=&Z(\lambda_i)^{-1}\sum_{n}\langle n(\lambda_i)|\mathcal{U}_{0,\tau}^{\dagger}\mathcal{G}_{\tau}\mathcal{U}_{0,\tau}\mathcal{U}_{0,\tau}^{-1}e^{-\beta z\mathcal{H}(\tau)}\mathcal{U}_{0,\tau}e^{-\beta(1-z) \mathcal{H}(0)}|n(\lambda_i)\rangle,\nonumber\\
&=&Z(\lambda_i)^{-1}\sum_{n}\langle n(\lambda_i)|\mathcal{G}_{0}\mathcal{U}_{0,\tau}^{-1}e^{-\beta z\mathcal{H}(\tau)}\mathcal{U}_{0,\tau}e^{-\beta(1-z) \mathcal{H}(0)}|n(\lambda_i)\rangle,\\
&=&Z(\lambda_i)^{-1}\text{Tr}_{\mathcal{G}_0}[\mathcal{U}_{0,\tau}^{-1}e^{-\beta z\mathcal{H}(\tau)}\mathcal{U}_{0,\tau}e^{-\beta(1-z) \mathcal{H}(0)}],\\
&=&Z(\lambda_i)^{-1}\text{Tr}_{\mathcal{G}_0}[e^{-\beta z\mathcal{H}(\tau)}\mathcal{U}_{0,\tau}e^{-\beta(1-z) \mathcal{H}(0)}\mathcal{U}_{0,\tau}^{-1}],\\
&=&Z(\lambda_i)^{-1}\text{Tr}_{\mathcal{G}_0}[\left(e^{-\beta\mathcal{H}(\tau)}\right)^z\left(\mathcal{U}_{0,\tau}e^{-\beta\mathcal{H}(0)}\mathcal{U}_{0,\tau}^{-1}\right)^{1-z}],\\
&=&\frac{Z(\lambda_f)^z}{Z(\lambda_i)^z}\text{Tr}_{\mathcal{G}_0}[\left(\Theta^{-1}\rho_R(0)\Theta\right)^z\left(\mathcal{U}_{0,\tau}\rho(0)\mathcal{U}_{0,\tau}^{-1}\right)^{1-z}],\\
&=&\frac{Z(\lambda_f)^z}{Z(\lambda_i)^z}\text{Tr}_{\mathcal{G}_0}[\left(\Theta^{-1}\rho_R(0)\Theta\right)^z\rho(\tau)^{1-z}],\\
&=&\frac{Z(\lambda_f)^z}{Z(\lambda_i)^z}\text{Tr}_{\mathcal{G}_0}[\left(\Theta^{-1}\rho_R(\tau-t)\Theta\right)^z\rho(t)^{1-z}],\\
&=&\exp\left[-\beta z\Delta F+(z-1)S_z\left(\Theta^{-1}\rho_R(\tau-t)\Theta||\rho(t)\right)\right].
\end{eqnarray}
Here $z$ is a finite real number and $\Delta F\equiv F(\beta,\lambda_f)-F(\beta,\lambda_i)$ is the free energy difference between the equilibrium states at the initial and final control parameters. In the last step, we have made use of definition of the order-$z$ quantum R\'{e}nyi divergences between two density matrices $\rho_1$ and $\rho_2$ \cite{Beigi2013,Lennert2013}, $S_z(\rho_1||\rho_2)\equiv\frac{1}{z-1}\ln[\text{Tr}[\rho_1^{z}\rho_2^{1-z}]]$, which is information theoretic generalization of standard relative entropy \cite{Renyi1961}. If we define $W-\Delta F$ as the dissipated work $W_{\text{diss}}$, we therefore obtain for quantum system with $\mathcal{PT}$ symmetry,
\begin{eqnarray}\label{central0}
\langle\Big( e^{-\beta W_{\text{diss}}}\Big)^z\rangle&=&\exp\left[(z-1)S_z\left(\Theta^{-1}\rho_R(\tau-t)\Theta||\rho(t)\right)\right].
\end{eqnarray}
Now we make several remarks on the relation:\\
(1). In Equation \eqref{central0}, $z$ is a free parameter. If we set $z=1$, Equation \eqref{central0} recovers the Jarzynski equality for $\mathcal{PT}$-symmetric quantum mechanics \cite{PT2015}.\\
(2). Fluctuation of the dissipated work in quantum system with $\mathcal{PT}$ symmetry is independent of time $t$ because the density operators appeared on the right hand side of Equation \eqref{central0} can be evaluated at any intermediate time $t\in[0,\tau]$.  This time independence is a consequence of the time reversal symmetry and unitarity of quantum dynamics. \\
(3). The quantum R\'{e}nyi divergences of order $z$ with $0<z<1$ is a valid measure of distinguishability \cite{Erven2014,Beigi2013} and therefore the R\'{e}nyi divergence appears in Equation \eqref{central0} for quantum system with $\mathcal{PT}$ symmetry is a quantification of the breaking of time reversal symmetry between the forward and its time reversed dynamics.
\\(4). Equation \eqref{central0} links the fluctuations of the dissipated work to the quantum R\'{e}nyi divergences between two non-equilibrium quantum states for $\mathcal{PT}$-symmetric quantum systems along the forward process and its time reversed process. It is clear that the dissipated work is a macroscopic quantity while on the other hand the Renyi divergences between two quantum states is a microscopic quantity. Moments of the dissipated work for quantum system with $\mathcal{PT}$ symmetry under time-dependent driving are given by,
\begin{eqnarray}
\langle W_{\text{diss}}^n\rangle&=&T^n\text{Tr}\Big[\rho(t)\mathcal{T}_n\Big(\ln[\rho(t)]-\ln[\Theta^{-1}\rho_R(\tau-t)\Theta]\Big)^n\Big],\label{momentsquan}\nonumber\\
\end{eqnarray}
where $T$ is the temperature, $n=1,2,3,\cdots$ and $\mathcal{T}_n$ is an ordering operator which sorts that in each term of the binomial expansion of $\Big(\ln[\rho(t)]-\ln[\Theta^{-1}\rho_R(\tau-t)\Theta]\Big)^n$, $\ln[\rho(t)]$ always sits on the left of $\ln[\Theta^{-1}\rho_R(\tau-t)\Theta]$.
In particular for $n=1$, it is
\begin{eqnarray}\label{ex}
\langle W_{\text{diss}}\rangle&=&TD[\rho(t)||\Theta^{-1}\rho_R(\tau-t)\Theta].\label{qdiss}
\end{eqnarray}
Here $D[\rho_1||\rho_2]\equiv \text{Tr}[\rho_1(\ln\rho_1-\ln\rho_2)]$ is the von Neumann relative entropy \cite{Wehrl1978} between two density matrices. Note that it was known that the mean value of dissipated work is related to the relative entropy between the forward dynamics and its time reversed dynamics for non-equilibrium dynamics in Hermitian systems \cite{Kawai2007,Jarzynski2006,Jarzynski2009,Parrondo2009,Deffner2010,Vedral2012,Serra2015}. \\
(5). Equation \eqref{central0} establishes an exact relation between the generating function of the dissipated work and the quantum R\'{e}nyi divergences between two non-equilibrium states and thus we can measure the family of R\'{e}nyi divergences from the non-equilibrium work measurement in a driving $\mathcal{PT}$-symmetric quantum system. Because the characteristic function of quantum work distribution can be measured from the Ramsey interference of a single spin \cite{Dorner2013,Mazzola2013,measure1}, one can also measure the family of R\'{e}nyi divergences between two quantum states in a $\mathcal{PT}$-symmetric quantum system.

If the Hamiltonian is in the regime of broken $\mathcal{PT}$ symmetry, the eigen spectrum has real and imaginary parts and thus the quantum dynamics is non-unitary. Since our derivations depend crucially on the unitarity of dynamics, we make conclude that the relation between dissipated work and Renyi divergences does not hold in the regime of broken $\mathcal{PT}$ symmetry.

\section{Physical Model Study}
Here we use a specific model to verify our central result, Equation \eqref{central0} in $\mathcal{PT}$-symmetric systems. Consider a two-level atom with the Hamiltonian \cite{PT2016,PTmodel1,PTmodel2}
\begin{eqnarray}
\mathcal{H}(t)&=&i\lambda(t)\sigma_z+J\sigma_x.
\end{eqnarray}
Here $\lambda(t)$ is a time-dependent control parameter and $J$ is a real parameter and $\sigma_{\alpha}$ is the Pauli operator in the $\alpha=x,y,z$-direction. This simple system has been realized experimentally in optics \cite{PTmodel1} and microcavities \cite{PTmodel2}.  The instantaneous eigen energies of the Hamiltonian are $E_{\pm}=\pm\sqrt{J^2-\lambda(t)^2}$ and thus the eigen energies are real if $|\lambda(t)|<J$ and they become purely imaginary if $|\lambda(t)|>J$.

We consider a sudden quench process, where the control parameter $\lambda$ is changed from the initial value $\lambda_i=0$ to a final value $\lambda_f=J/2$. And their corresponding Hamiltonians are respectively
\begin{eqnarray}
\mathcal{H}_i&=&J\sigma_x,\\
\mathcal{H}_f&=&\frac{iJ}{2}\sigma_z+J\sigma_x.
\end{eqnarray}
In these parameter regime, the system is in the $\mathcal{PT}$ symmetry unbroken phase and thus our central results should be valid in this process. We are now proving this is true. First of all, the generating work of work in this sudden quench process is
\begin{eqnarray}
G(u)&=&\int dWP(W)e^{iuW},\\
&=&Z_i^{-1}\text{Tr}_{\mathcal{G}_i}[e^{-(\beta+iu)\mathcal{H}_i}e^{iu\mathcal{H}_f}],\\
&=&\frac{\cosh(\eta)\cos(\kappa)-i\frac{2}{\sqrt{3}}\sinh(\eta)\sin(\kappa)}{2\cosh(\beta J)}.
\end{eqnarray}
Here we define $\eta\equiv(\beta+iu)J$ and $\kappa\equiv\sqrt{3}Ju/2$.

The quantum R\'{e}nyi divergences between the initial state $\rho_i$ and the equilibrium state at the final control parameter $\rho_f$ is
\begin{eqnarray}
&&S_z(\rho_f||\rho_i)=\frac{1}{z-1}\ln\left(\text{Tr}_{\mathcal{G}_f}[\rho_f^z\rho_i^{1-z}]\right)\\
&=&\frac{1}{z-1}\ln\left(\frac{\cosh(\theta)\cosh(\delta)+\frac{2}{\sqrt{3}}\sinh(\theta)\sinh(\delta)}{[2\cosh(\beta J)]^{1-z}[2\cosh(\frac{\sqrt{3}\beta J}{2})]^{z}}\right).
\end{eqnarray}
Here we define $\theta\equiv\beta(1-z)J$ and $\delta\equiv\sqrt{3}z\beta J/2$. In addition, the free energy difference between the equilibrium states at final control parameter and the initial control parameter is
\begin{eqnarray}
e^{-\beta \Delta F}&=&e^{-\beta [F(\beta,\lambda_f)-F(\beta,\lambda_i)]}=\frac{\cosh(\frac{\sqrt{3}}{2}\beta J)}{\cosh(\beta J)}.
\end{eqnarray}
Then it is easy to check that in this sudden quench process,
\begin{eqnarray}
\langle\left(e^{-\beta W}\right)^z\rangle&=&G(i\beta z)=e^{-\beta z\Delta F}e^{(z-1)S_z(\rho_f||\rho_i)}.
\end{eqnarray}
Thus the relation between the dissipated work and the quantum Renyi divergences in $\mathcal{PT}$-symmetric system is valid.

\section{Conclusions}  We have shown that the relations between the dissipated work done on a system driven arbitrarily far from equilibrium and the quantum R\'{e}nyi divergences generalized to the $\mathcal{PT}$-symmetric quantum systems. We have found that for quantum systems in the regime of unbroken $\mathcal{PT}$ symmetry, the relation between dissipation and R\'{e}nyi divergences holds, while on the other hand the relation is not valid for the quantum system in the $\mathcal{PT}$ symmetry broken regime. These findings have been demonstrated in an experimentally
relevant system and could be experimentally verified in the setup consisting of two coupled cavities.

\begin{acknowledgements}
This work was supported by National Natural Science Foundation of China (Grants No. 11604220) and the Start Up Fund of Shenzhen University (Grants No. 2016018).
\end{acknowledgements}

\subsection*{Appendix A: Unitarity Condition in $\mathcal{PT}$-symmetric Quantum Mechanics}
 \renewcommand{\theequation}{A\arabic{equation}} \setcounter{equation}{0}
In this appendix, we prove that unitarity condition Equation \eqref{unitarity} in the main text. Consider two different initial states, $|\phi_1(0)\rangle$ and $|\phi_2(0)$ and they evolve in time under the same Hamiltonian $\mathcal{H}(t)$ and the corresponding time evolution operator is denoted as $\mathcal{U}_{0,t}$. Then the time evolved states at time $t$ are respectively $|\phi_1(t)\rangle=\mathcal{U}_{0,t}|\phi_1(0)\rangle$ and $|\phi_2(t)\rangle=\mathcal{U}_{0,t}|\phi_2(0)\rangle$. Unitarity means that
\begin{eqnarray}
\langle\phi_1(0)|\mathcal{G}_0|\phi_2(0)\rangle=\langle\phi_1(t)|\mathcal{G}_t|\phi_2(t)\rangle.
\end{eqnarray}
While the right hand side of the above equation can be written as,
\begin{eqnarray}
\langle\phi_1(t)|\mathcal{G}_t|\phi_2(t)\rangle&=&\langle\phi_1(0)|\mathcal{U}_{0,t}^{\dagger}\mathcal{G}_t\mathcal{U}_{0,t}|\phi_2(0)\rangle.
\end{eqnarray}
Thus we have
\begin{eqnarray}
\langle\phi_1(0)|\mathcal{G}_0|\phi_2(0)\rangle=\langle\phi_1(0)|\mathcal{U}_{0,t}^{\dagger}\mathcal{G}_t\mathcal{U}_{0,t}|\phi_2(0)\rangle.
\end{eqnarray}
This equality is valid for any non-zero states $|\phi_1(0)\rangle$ and $|\phi_2(0)\rangle$ and we thus derived Equation \eqref{unitarity} in the main text.

\subsection*{Appendix B: Time Evolution of A Density Operator in $\mathcal{PT}$-symmetric Quantum Mechanics}
 \renewcommand{\theequation}{B\arabic{equation}} \setcounter{equation}{0}
In this appendix, we prove that time evolution of an arbitrary density operator, Equation \eqref{ef} in the main text. Let us consider an arbitrary nonzero density operator $\rho(0)$, normalization requires that
\begin{eqnarray}
\text{Tr}_{\mathcal{G}_0}[\rho(0)]=1.
\end{eqnarray}
Making use of a complete basis states $\{|\phi_n(0)\rangle\}$, the trace operation can be written as
\begin{eqnarray}
\text{Tr}_{\mathcal{G}_0}[\rho(0)]=\sum_n\langle\phi_n(0)|\mathcal{G}_0\rho(0)|\phi_n(0)\rangle=1.
\end{eqnarray}
We assume that the density operator $\rho(0)$ is evolved into $\rho(t)$ in time $t$. Then the normalization requires that
\begin{eqnarray}
\text{Tr}_{\mathcal{G}_t}[\rho(t)]=1.
\end{eqnarray}
Making use of the complete basis states $\{|\phi_n(t)\rangle=\mathcal{U}_{0,t}|\phi_n(0)\rangle\}$, the trace operation can be written as
\begin{eqnarray}
\text{Tr}_{\mathcal{G}_t}[\rho(t)]=\sum_n\langle\phi_n(0)|\mathcal{U}_{0,t}^{\dagger}\mathcal{G}_t\rho(t)\mathcal{U}_{0,t}|\phi_n(0)\rangle=1.
\end{eqnarray}
Thus we have
\begin{eqnarray}
\sum_n\langle\phi_n(0)|\mathcal{G}_0\rho(0)|\phi_n(0)\rangle=\sum_n\langle\phi_n(0)|\mathcal{U}_{0,t}^{\dagger}\mathcal{G}_t\rho(t)\mathcal{U}_{0,t}|\phi_n(0)\rangle.\nonumber\\
\end{eqnarray}
This equality is valid for any orthonormal basis states $\{|\phi_n(0)\rangle\}$ and we thus get
\begin{eqnarray}
\mathcal{G}_0\rho(0)=\mathcal{U}_{0,t}^{\dagger}\mathcal{G}_t\rho(t)\mathcal{U}_{0,t}.
\end{eqnarray}
Making use of the unitarity condition, Equation \eqref{unitarity}, we thus get
\begin{eqnarray}
\rho(t)&=&\mathcal{G}_t^{-1}\left(\mathcal{U}_{0,t}^{\dagger}\right)^{-1}\mathcal{G}_0\rho(0)\mathcal{U}_{0,t}^{-1},\\
&=&\mathcal{U}_{0,t}\rho(0)\mathcal{U}_{0,t}^{-1}.
\end{eqnarray}
Thus Equation \eqref{ef} in the main text is proved.

\subsection*{Appendix C: Relation between Time Evolution Operator in a Non-equilibrium Process and In its Time Reversed Process for $\mathcal{PT}$-symmetric Quantum Mechanics}
 \renewcommand{\theequation}{C\arabic{equation}} \setcounter{equation}{0}
In this appendix, we prove the relation between time evolution operators in the forward process and that in its time reversed process, Equation \eqref{frrelation}.
In the forward process, we have
\begin{eqnarray}\label{fo}
\rho(\tau)=\mathcal{U}_{\tau-t,\tau}\rho(\tau-t)U_{\tau-t,\tau}^{-1}.
\end{eqnarray}
Now if we make a time reversal operation on the final state $\rho(\tau)$ and we get $\Theta\rho(\tau)\Theta^{-1}$ and then we drive this time reversed state by the evolution operator in the time reversed process, $\mathcal{V}_{0,t}$, for a time duration $t$. According to time reversal symmetry, the final evolved state is $\Theta\rho(\tau-t)\Theta^{-1}$. This gives us
\begin{eqnarray}
\Theta\rho(\tau-t)\Theta^{-1}=\mathcal{V}_{0,t}\Theta\rho(\tau)\Theta^{-1}\mathcal{V}_{0,t}^{-1}.
\end{eqnarray}
Making use of Equation \eqref{fo}, we get
\begin{eqnarray}
\Theta\rho(\tau-t)\Theta^{-1}=\mathcal{V}_{0,t}\Theta\mathcal{U}_{\tau-t,\tau}\rho(\tau-t)U_{\tau-t,\tau}^{-1}\Theta^{-1}\mathcal{V}_{0,t}^{-1}.
\end{eqnarray}
Thus we obtain
\begin{eqnarray}
\Theta=\mathcal{V}_{0,t}\Theta\mathcal{U}_{\tau-t,\tau}.
\end{eqnarray}
This leads to,
\begin{eqnarray}
\mathcal{V}_{0,t}=\Theta\mathcal{U}_{\tau-t,\tau}^{-1}\Theta^{-1}.
\end{eqnarray}
Therefore Equation \eqref{frrelation} in the main text is proved.

\end{document}